\documentclass[12pt]{article}
\usepackage{oldgerm}
\usepackage{yfonts}
\usepackage{euler}
\usepackage{graphics}
\usepackage{bm}
\usepackage{graphicx}
\usepackage{amsmath}
\usepackage{amssymb}
\usepackage{amstext}
\usepackage{epstopdf}
\usepackage{amscd}
\usepackage{amsfonts}
\usepackage{appendix}
\usepackage{color}
\usepackage{wasysym}
\DeclareMathAlphabet{\EuFrak}{U}{euf}{m}{n}
\DeclareMathAlphabet{\EuScript}{U}{eus}{m}{n}

\newcommand{\nd}{\noindent}

\newcommand{\be}{\begin{equation}}
\newcommand{\ee}{\end{equation}}
\newcommand{\ben}{\begin{eqnarray}}
\newcommand{\een}{\end{eqnarray}}

\title{{\bf Hidden correlations entailed by  q-non additivity  
render  the q-monoatomic gas highly non trivial}}

\author{{A. Plastino$^1$, M. C. Rocca$^1$}\\, 
\small{$^1$ La Plata National University and
Argentina's National Research Council}\\
\small{(IFLP-CCT-CONICET)-C. C. 727, 1900 La Plata - Argentina}}

\date{\today}

\begin{document}

\maketitle

\begin{abstract}

\nd It ts known that Tsallis' q-non-additivity entails hidden
correlations. It has also been shown that even for a monoatomic  gas,
both the q-partition function $Z$ and the mean energy $<U>$
diverge and, in particular, exhibit poles for certain values of the 
Tsallis non additivity parameter $q$. 
This happens because $Z$ and $<U>$ both depend on a 
$\Gamma$-function. This $\Gamma$, in turn, depends upon the spatial
dimension $\nu$. We encounter three different regimes
according to the argument $A$ of the $\Gamma$-function.
(1) $A>0$, (2) $A<0$ and $\Gamma>0$ outside the poles.
(3) $A$ displays poles and the physics is obtained via
 dimensional regularization.
  In cases (2) and (3) 
one discovers gravitational effects and quartets of particles.
Moreover, bound states and gravitational effects 
emerge as a consequence of the hidden q-correlations.

\nd Keywords: q-Statistics, divergences,  partition function,
dimensional regularization, specific heat.

\end{abstract}

\newpage

\renewcommand{\theequation}{\arabic{section}.\arabic{equation}}

\section{Introduction}

\setcounter{equation}{0}

\nd Generalized or q-statistical mechanics \`a la Tsallis has
generated manifold applications in the last quarter of a century 
\cite{tsallis,web,epjb1,epjb2,epjb3,epjb4,epjb5,epjb6,
epjb7,epjb8,epjb9,li1,li2}. It has been shown (see for instance,
\cite{PP93,chava}) that  Tsallis' q-statistics is of great
importance for dealing with some astrophysical issues involving
self-gravitating systems \cite{lb}. Moreover, this statistics has
proved its utility in many different scientific fields, with several
thousands  of  publications and authors \cite{web}, so that studying
its structural features is an important issue for physics,
astronomy, biology, neurology, economics, etc. \cite{tsallis}. The
success of the q-statistics reaffirms the well grounded notion
asserting that there is much physics whose origin is of purely
statistical nature (not mechanical). As a spectacular example, me
mention the application of q-ideas to high energy experimental
physics, where the q-statistics appears to adequately describe the
transverse momentum distributions of different hadrons
\cite{tp11o,tp11,phenix}. 

\nd In this work we show that as yet unexplored  gravitational
effects characterize this q-theory on account of {\it divergences}
that, in some circumstances, emerge, within the q-statistical
framework,  in both the mean energy $<U>$   and the partition 
function $Z$ \cite{epjb}.

\nd Divergences constitute an important issue in theoretical physics.
Indeed, the study and elimination of divergences of a physical
theory is perhaps one of the most important aspects of theoretical
work. The quintessential typical example is the thus far failed, attempt to
quantify the gravitational field. Some examples of elimination of divergences can be looked at 
in  references \cite{tq1,tq2,tq3,tq4,tq5}.

\nd  We will use here  an extremely  simplified version, such as that of \cite{tr1}, of the
ideas of \cite{tq1,tq2,tq3,tq4,tq5} in connection with  Tsallis
q-statistics \cite{tsallis,web}, with emphasis in its
applicability to gravitational  issues \cite{PP93,chava}, in
particular  self-gravitating systems \cite{lb}.
We will see that the removal of the above mentioned divergences
produces interesting insights.

\nd It is to be stressed that for $Z<0$ the system can not
exist, as no probabilities can be introduced. Then, several
options are available.
$<U>>0$ is the natural state of
affairs. We will uncover that, in a Tsallis' scenario, $<U><0$
is possible for special q-values which entails boundedness.
Also negative specific heats $C$ may emerge, an indication of
gravitational effects \cite{lb}.

\nd These interesting results appear after using mathematics
known since at least 40 years ago and for whose development
M. Veltman and G. t'Hooft where awarded the
Nobel prize of physics in 1999.
Acquaintance with such
mathematics is not really needed to understand this paper. 
The reader has merely to accept that their physical significance 
can not be doubted. In a few words, one needs from the above mathematics analytical
extensions and dimensional regularization \cite{tq1,tq2,tq3,tq4,tq5}. 

\nd In this work we analyze the behavior of
$Z$ and $<U>$ in three regions of the argument of the 
$\Gamma$-function contained in them.
The nature of the argument of the $\Gamma$-function
governs the behavior of $Z$ and $<U>$.
This behavior yields three different regions,
for a given spatial dimension $\nu$, Tsallis' index
$q$ and number of particles $N$.
The region's particularities are:\\
$(1)\;\;\;\frac {1} {1-q}-\frac {n\nu} {2}-1>0$\\ 
$(2)\;\;\;\frac {1} {1-q}-\frac {n\nu} {2}<0\;\;\;
\Gamma\left(\frac {1} {1-q}-\frac {n\nu} {2}\right)>0$\\
$(3)\;\;\;\frac {1} {1-q}-\frac {n\nu} {2}=-p
\;\;\;p=0,1,2,3,4.....$\\ 
Normal monoatomic gas behavior is
encountered in region (1). Instead, gravitational
effects are discovered in region (2).
Finally, in region (3) we have both normal
behavior and gravitational effects.
In particular the $N$ particles are grouped
into quartets that remind one of alpha particles.
\nd {\bf It should be noted than in case (3) we are making a
regularization of the corresponding theory and NOT
a renormalization.}

\section{The monoatomic gas}

\setcounter{equation}{0}
We stress that here  we use
normal (linear in the probability) expectation values, and not the weighted
ones usually attached to Tsallis' theory \cite{tsallis}. 
This is done for simplicity. Other ways of evaluating mean values
pose difficulties in this context that will be tackled in the future
\cite{referee}.

Remark  that in this case, restricting  ourselves to 
the interval $[0< q \le 1]$, the so-called Tsallis cut-off  \cite{tsallis} does not apply. 

The q-partition function of a monoatomic gas is given by:
\begin{equation}
\label{eq0.1}
Z=V^n\int\limits_{-\infty}^{\infty}
\left[1+\frac {\beta(1-q)} {2m}(p_1^2+p_2^2+\cdot\cdot\cdot
p_n^2)\right]^{\frac {1}{q-1}}d^{\nu}p_1d^{\nu}p_2\cdot\cdot
\cdot d^{\nu}p_n,
\end{equation}
The mean energy is defined by:
\[<U>=\frac {V^n} {Z}\int\limits_{-\infty}^{\infty}
\left[1+\frac {\beta(1-q)} {2m}(p_1^2+p_2^2+\cdot\cdot\cdot
p_n^2)\right]^{\frac {1}{q-1}}\]
\begin{equation}
\label{eq0.2}
\frac {p_1^2+p_2^2+\cdot\cdot\cdot p_n^2} {2m}
d^{\nu}p_1d^{\nu}p_2\cdot\cdot
\cdot d^{\nu}p_n.
\end{equation}

Both integrals are in general divergent for many values
of $q$.  This can be
proved by a simple powers count.  We appeal to techniques developed in \cite{tq1,tq2,tq3,tq4,tq5} so as to deal with them.

In  ref.\cite{tr1} we have computed both  $Z$ and $<U>$
obtaining:
\begin{equation}
\label{eq2.1} 
Z=V^n\left[\frac {2\pi m} {\beta(1-q)}\right]^{
\frac {\nu n} {2}}
\frac {\Gamma\left(\frac {1} {1-q}-\frac {\nu n} {2}\right)}
{\Gamma\left(\frac {1} {1-q}\right)}.
\end{equation}
\begin{equation}
\label{eq2.2} 
<U>=\frac {V^n} {Z}\frac {\nu n} {2\beta(1-q)}
\left[\frac {2\pi m} {\beta(1-q)}\right]^{\frac {\nu n} {2}}
\frac {\Gamma\left(\frac {1} {1-q}-\frac {\nu n} {2}
-1\right)} {\Gamma\left(\frac {1} {1-q}\right)}.
\end{equation}
\begin{equation}
\label{eq2.3} 
<U>=\frac {\nu n} {\beta[2q-\nu n(1-q)]}.
\end{equation}
The derivative with respect to  $T$ yields for the specific heat 
at constant  volume

\begin{equation}
\label{eq2.4} 
C_V=\frac {\nu nk} {2q-\nu n(1-q)}.
\end{equation}

\section{Limitations for the particle-number}

\setcounter{equation}{0}

We briefly review now results obtained in Ref. \cite{tr1}. Some related work by  Livadiotis,  McComas, and Obregon,  should be cited 
\cite{li1,li2,obreg}.  We pass first to analyze the Gamma functions 
involved in computing $Z$ and $<U>$, for the region $[0 < q \le 1]$. 
From  (\ref{eq2.1}) we have, for positive Gamma-argument 
\begin{equation}
\label{eq3.1} 
\frac {1} {1-q}-\frac {\nu n} {2}>0.
\end{equation}
In the same vein we have from  (\ref{eq2.2}) 
\begin{equation}
\label{eq3.2} 
\frac {1} {1-q}-\frac {\nu n} {2}-1>0.
\end{equation}
We reach then  two conditions that  pose severe limitations on the 
particle-number $n$, i.e., 
\begin{equation}
\label{eq3.3} 
1\leq n<\frac {2q} {\nu(1-q)}
\end{equation}
There is a maximum permissible $n$.
For example, if 
$q=1-10^{-3}, \nu=3$, one has
\begin{equation}
\label{eq3.4} 
1\leq n<666.
\end{equation}
We can not have more than 665 particles. Keeping the dimensionality equal to three, for $q=1/2$ just one particle is permitted and for $q=1/4$, no particles can be present exist. Roughly, for  a number of particles of the order of $10^n$,
$q$ has to be  of the order of $1-10^{-n}$.

\section{The dimensional  analytical extension of divergent integrals \cite{tq1,tq2,tq3,tq4,tq5}}

\setcounter{equation}{0}

The exposition of our present results starts here. We pass first to considering negative Gamma arguments in (\ref{eq2.1}), which will require analytical extension/dimensional regularization  in integrals (2.1) and (2.2). One has

\begin{equation}
\label{eq4.1}
\frac {1} {1-q}-\frac {\nu n} {2}<0,
\end{equation}
together with
\begin{equation}
\label{eq4.2}
\Gamma\left(\frac {1} {1-q}-\frac {\nu n} {2}\right)>0,
\end{equation}
We use now
\begin{equation}
\label{eq4.3}
\Gamma(z)\Gamma(1-z)=\frac {\pi} {\sin(\pi z)},
\end{equation}
to find
\begin{equation}
\label{eq4.4}
\Gamma\left(\frac {1} {1-q}-\frac {\nu n} {2}\right)=-
\frac {\pi} {\sin\pi\left(\frac {\nu n} {2}-\frac {1} {1-q}\right)
\Gamma\left(\frac {\nu n} {2}-\frac {1} {1-q}\right)}>0.
\end{equation}
This is true if 
\begin{equation}
\label{eq4.5}
\sin\pi\left(\frac {\nu n} {2}-\frac {1} {1-q}\right)<0,
\end{equation}
so that 
\begin{equation}
\label{eq4.6}
2p+1<\frac {\nu n} {2}-\frac {1} {1-q}<2(p+1)
\end{equation}
where $p=0,1,2,3,4,5.....$, or equivalently

\begin{equation}
\label{eq4.7}
\frac {4p+2} {\nu}+\frac {2} {\nu(1-q)}<n<
\frac {4(p+1)} {\nu}+\frac {2} {\nu(1-q)}.
\end{equation}
Now, from  (\ref{eq2.1}), (\ref{eq2.3}), and  (\ref{eq2.4})
we obtain (a) $Z>0$, (b) $<U><0$, (c) $C<0$, 
which entails bound states, on account of (b) and self-gravitation according to (c) \cite{lb}.

\section{The poles of the q-Ideal Gas treatment}

\setcounter{equation}{0}

If the Gamma's argument is such that 

\begin{equation}
\label{eq5.1}
\frac {1} {1-q}-\frac {\nu n} {2}=-p\;\;{\rm for} \;\;p=0,1,2,3,......,
\end{equation}
 $Z$ displays a single pole. \vskip 3mm

\nd For $\nu=1$ we have
\begin{equation}
\label{eq5.2}
\frac {1} {1-q}-\frac {n} {2}=-p\;\;{\rm for} \;\;p=0,1,2,3,.......
\end{equation}
\nd Since  $0\leq q<1$, the concomitant 
 $q$ values are 
\begin{equation}
\label{eq5.3}
q=\frac {1} {2},\frac {2} {3},\frac {3} {4},\frac {4} {5},......,
\end{equation}
$n$ even, $n\geq 4$ and

\begin{equation}
\label{eq5.4}
q=\frac {1} {3},\frac {3} {5},\frac {5} {7},\frac {7} {9},......,
\end{equation}
$n$ odd, $n\geq 3$.

\nd For $\nu=2$
\begin{equation}
\label{eq5.5}
\frac {1} {1-q}-n=-p\;\;{\rm for} \;\;p=0,1,2,3,......,
\end{equation}
\nd Again, since  $0\leq q<1$, 
\begin{equation}
\label{eq5.6}
q=\frac {1} {2},\frac {2} {3},\frac {3} {4},\frac {4} {5},......,
\end{equation}
$n\geq 2$.

\nd For $\nu=3$
\begin{equation}
\label{eq5.7}
\frac {1} {1-q}-\frac {3n} {2}=-p\;\;{\rm for} \;\;p=0,1,2,3,......,
\end{equation}
and because  $0\leq q<1$,
\begin{equation}
\label{eq5.8}
q=\frac {1} {2},\frac {2} {3},\frac {3} {4},\frac {4} {5},......,
\end{equation}
$n$ even, $n\geq 1$, and
\begin{equation}
\label{eq5.9}
q=\frac {1} {3},\frac {3} {5},\frac {5} {7},\frac {7} {9},......,
\end{equation}
$n$ odd, $n\geq 1$.

The location of these poles will change if
escort mean values where used.

\nd We discuss now poles in $<U>$, given by 

\begin{equation}
\label{eq5.10}
\frac {1} {1-q}-\frac {\nu n} {2}-1=-p\;\;{\rm for} \;\;p=0,1,2,3,......,
\end{equation}
For $\nu=1$
\begin{equation}
\label{eq5.11}
\frac {1} {1-q}-\frac {n} {2}-1=-p\;\;{\rm for} \;\;p=0,1,2,3,......,
\end{equation}
On account on the condition (G): $0\leq q<1$ we have
\begin{equation}
\label{eq5.12}
q=\frac {1} {2},\frac {2} {3},\frac {3} {4},\frac {4} {5},......,
\end{equation}
for $n$ even, $n\geq 2$ and
\begin{equation}
\label{eq5.13}
q=\frac {1} {3},\frac {3} {5},\frac {5} {7},\frac {7} {9},......,
\end{equation}
for $n$ odd , $n\geq 1$.

\nd For $\nu=2$
\begin{equation}
\label{eq5.14}
\frac {1} {1-q}-n-1=-p\;\;{\rm for} \;\;p=0,1,2,3,......,
\end{equation}
minding (G) we have

\begin{equation}
\label{eq5.15}
q=\frac {1} {2},\frac {2} {3},\frac {3} {4},\frac {4} {5},......,
\end{equation}
$n\geq 1$.

\nd For $\nu=3$
\begin{equation}
\label{eq5.16}
\frac {1} {1-q}-\frac {3n} {2}-1=-p\;\;{\rm for} \;\;p=0,1,2,3,......,
\end{equation}
and, from (G), 
\begin{equation}
\label{eq5.17}
q=\frac {1} {2},\frac {2} {3},\frac {3} {4},\frac {4} {5},......,
\end{equation}
$n$ even, $n\geq 1$ and
\begin{equation}
\label{eq5.18}
q=\frac {1} {3},\frac {3} {5},\frac {5} {7},\frac {7} {9},......,
\end{equation}
$n$ odd, $n\geq 1$.

\section{The three-dimensional case}

\setcounter{equation}{0}

As an example of dimensional regularization  \cite{tq1,tq2,tq3,tq4,tq5}
we will go into some detail concerning the poles at  $q=\frac {1} {2}$ and $q=\frac {1} {3}$.

\subsection{The $q=1/2$ pole}

In this case $n$is  even, $n\geq 2$.
We have
\begin{equation}
\label{eq6.1} 
Z=V^n\left(\frac {4m\pi} {\beta}\right)^{\frac {\nu n} {2}}
\Gamma\left(2-\frac {\nu n} {2}\right).
\end{equation}
Using
\begin{equation}
\label{eq6.2} 
\Gamma\left(2-\frac {\nu n} {2}\right)
\Gamma\left(\frac {\nu n} {2}-1\right)=
-\frac {\pi} {\sin\left(\frac {\pi\nu n} {2}\right)}
\end{equation}
or, equivalenly

\begin{equation}
\label{eq6.3} 
\Gamma\left(2-\frac {\nu n} {2}\right)
\Gamma\left(\frac {\nu n} {2}-1\right)=
-\frac {(-1)^{\frac {3n} {2}+1}\pi} {\sin\left[
\frac {\pi n} {2} (\nu-3)\right]},
\end{equation}
so that 
\begin{equation}
\label{eq6.4} 
Z=V^n\left(\frac {4m\pi} {\beta}\right)^{\frac {\nu n} {2}}
\frac {(-1)^{\frac {3n} {2}+1}\pi}
{\sin[\frac {\pi n} {2}(\nu-3)]
\Gamma\left(\frac {\nu n} {2}-1\right)}.
\end{equation}
Since 
\begin{equation}
\label{eq6.5} 
\sin[\frac {\pi n} {2}(\nu-3)]=\frac {\pi n} {2}(\nu-3)
\left\{1+\sum\limits_{m=1}^{\infty}
\frac {(-1)^m} {(2m+1)!} \left[\frac {\pi n} {2}
(\nu-3)\right]^{2m}\right\}=
\end{equation}
\begin{equation}
\label{eq6.6} 
=\frac {\pi n} {2}(\nu-3)X,
\end{equation}
with
\begin{equation}
\label{eq6.7} 
X=
\left\{1+\sum\limits_{m=1}^{\infty}
\frac {(-1)^m} {(2m+1)!} \left[\frac {\pi n} {2}
(\nu-3)\right]^{2m}\right\},
\end{equation}
we get
\begin{equation}
\label{eq6.8} 
Z=V^n\left(\frac {4m\pi} {\beta}\right)^{\frac {3 n} {2}}
\frac {(-1)^{\frac {3n} {2}+1}}
{\Gamma\left(\frac {\nu n} {2}-1\right)
X\frac {n} {2}(\nu-3)}
\left[1+\frac {n} {2}(\nu-3)\ln\left(\frac {4m\pi} {\beta}\right)+
\cdot\cdot\cdot\right]
\end{equation}
The term independent of  $\nu-3$ is, following dimensional regularization prescriptions 
\cite{tq1,tq2,tq3,tq4,tq5}
\begin{equation}
\label{eq6.9} 
Z=V^n\left(\frac {4m\pi} {\beta}\right)^{\frac {3 n} {2}}
\frac {(-1)^{\frac {3n} {2}+1}}
{\Gamma\left(\frac {3 n} {2}-1\right)}
\ln\left(\frac {4m\pi} {\beta}\right)
\end{equation}
This is then the physical Z-value at the pole \cite{tq1,tq2,tq3,tq4,tq5}. For the mean energy we have

\begin{equation}
\label{eq6.10} 
Z<U>=V^n\frac {n\nu} {\beta}
\left(\frac {4m\pi} {\beta}\right)^{\frac {\nu n} {2}}
\Gamma\left(1-\frac {\nu n} {2}\right).
\end{equation}
Using
\begin{equation}
\label{eq6.11} 
\Gamma\left(1-\frac {\nu n} {2}\right)
\Gamma\left(\frac {\nu n} {2}\right)=
\frac {\pi} {\sin\left(\frac {\pi\nu n} {2}\right)}
\end{equation}
or, equivalently
\begin{equation}
\label{eq6.12} 
\Gamma\left(1-\frac {\nu n} {2}\right)
\Gamma\left(\frac {\nu n} {2}\right)=
-\frac {(-1)^{\frac {3n} {2}}\pi} {\sin\left[
\frac {\pi n} {2} (\nu-3)\right]}
\end{equation}
one finds for  $<U>$

\begin{equation}
\label{eq6.13} 
Z<U>=V^n\frac {n\nu} {\beta}
\left(\frac {4m\pi} {\beta}\right)^{\frac {\nu n} {2}}
\frac {(-1)^{\frac {3n} {2}}\pi}
{\sin[\frac {\pi n} {2}(\nu-3)]
\Gamma\left(\frac {\nu n} {2}\right)}.
\end{equation}
$<U>$ can be recast as

\[Z<U>=V^n\frac {n(\nu-3)} {\beta}
\left(\frac {4m\pi} {\beta}\right)^{\frac {\nu n} {2}}
\frac {(-1)^{\frac {3n} {2}}\pi}
{\sin[\frac {\pi n} {2}(\nu-3)]
\Gamma\left(\frac {\nu n} {2}\right)} +\]
\begin{equation}
\label{eq6.14} 
V^n\frac {3n} {\beta}
\left(\frac {4m\pi} {\beta}\right)^{\frac {\nu n} {2}}
\frac {(-1)^{\frac {3n} {2}}\pi}
{\sin[\frac {\pi n} {2}(\nu-3)]
\Gamma\left(\frac {\nu n} {2}\right)}.
\end{equation}
 Repeating the Z-treatment yields for  $<U>$:

\[Z<U>=V^n\frac {2} {\beta}
\left(\frac {4m\pi} {\beta}\right)^{\frac {3 n} {2}}
\frac {(-1)^{\frac {3n} {2}}}
{\Gamma\left(\frac {3 n} {2}\right)}+\]
\begin{equation}
\label{eq6.15} 
V^n\frac {3n} {\beta}
\left(\frac {4m\pi} {\beta}\right)^{\frac {3 n} {2}}
\frac {(-1)^{\frac {3n} {2}}}
{\Gamma\left(\frac {3 n} {2}\right)}
\ln\left(\frac {4m\pi} {\beta}\right)
\end{equation}
or,  equivalently

\begin{equation}
\label{eq6.16} 
Z<U>=V^n\frac {2} {\beta}
\left(\frac {4m\pi} {\beta}\right)^{\frac {3 n} {2}}
\frac {(-1)^{\frac {3n} {2}}}
{\Gamma\left(\frac {3 n} {2}\right)}
\left[1-\frac {3n} {2}
\ln\left(\frac {4m\pi} {\beta}\right)\right].
\end{equation}
Appealing here to  (\ref{eq6.9}) for the physical $Z$ 
 we finally obtain

\begin{equation}
\label{eq6.17} 
<{\cal U}>=
\frac {4} {\beta(3n-2)}
\left[\frac {1} {\ln\beta-\ln 4m\pi}-
\frac {3n} {2}\right].
\end{equation}
We discuss first $(-1)^{\frac {3n} {2}+1}=-1$ and then   
$n=4,8,12,16......$, so that 

\begin{equation}
\label{eq6.18} 
Z=\frac {V^n}
{\Gamma\left(\frac {3 n} {2}-1\right)}
\left(\frac {4m\pi} {\beta}\right)^{\frac {3 n} {2}}
\ln\left(\frac {\beta} {4m\pi}\right)
\end{equation}
If
$(-1)^{\frac {3n} {2}+1}=1$, then  
$n=2,6,10,14......$ and
\begin{equation}
\label{eq6.19} 
Z=\frac {V^n}
{\Gamma\left(\frac {3 n} {2}-1\right)}
\left(\frac {4m\pi} {\beta}\right)^{\frac {3 n} {2}}
\ln\left(\frac {4m\pi} {\beta}\right).
\end{equation}
From (\ref{eq6.17}) - (\ref{eq6.18}) and requiring  
$Z>0$ and  $<U>>0$ one deduces
\begin{equation}
\label{eq6.20}
\frac {1} {4m\pi ke^{\frac {2} {3n}}}<T<
\frac {1} {4m\pi k}.
\end{equation}
From  (\ref{eq6.17}) -  (\ref{eq6.19}) and demanding  
$Z>0$ y $<U><0$ one finds 
\begin{equation}
\label{eq6.21}
0\leq T<\frac {1} {4m\pi ke^{\frac {2} {3n}}}
\end{equation}
The specific heat derives from 
(\ref{eq6.17}) for $<U>$. One has
\begin{equation}
\label{eq6.22} 
C=
\frac {4k} {3n-2}
\left[\frac {1} {\ln\beta-\ln 4m\pi}+
\frac {1} {(\ln\beta-\ln 4m\pi)^2}-
\frac {3n} {2}\right].
\end{equation}

\subsection{The $q=1/3$ pole}

Here  $Z$ is
\begin{equation}
\label{eq6.23} 
Z=V^n\left(\frac {3m\pi} {\beta}\right)^{\frac {\nu n} {2}}
\frac {\Gamma\left(\frac {3} {2}-\frac {\nu n} {2}\right)}
{\Gamma\left(\frac {3} {2}\right)}.
\end{equation}
Using again
\begin{equation}
\label{eq6.24} 
\Gamma\left(\frac {3} {2}-\frac {\nu n} {2}\right)
\Gamma\left(\frac {\nu n} {2}-\frac {1} {2}\right)=
-\frac {\pi} {\cos\left(\frac {\pi\nu n} {2}\right)},
\end{equation}
or, equivalently

\begin{equation}
\label{eq6.25} 
\Gamma\left(\frac {3} {2}-\frac {\nu n} {2}\right)
\Gamma\left(\frac {\nu n} {2}-\frac {1} {2}\right)=
-\frac {(-1)^{\frac {3n-1} {2}}\pi} {\sin\left[
\frac {\pi n} {2} (\nu-3)\right]}, 
\end{equation}
so that 
\begin{equation}
\label{eq6.26} 
Z=\frac {V^n} {\Gamma\left(\frac {3} {2}\right)}
\left(\frac {3m\pi} {\beta}\right)^{\frac {\nu n} {2}}
\frac {(-1)^{\frac {3n-1} {2}}\pi}
{\sin[\frac {\pi} {2}(\nu-3)]
\Gamma\left(\frac {\nu n} {2}-\frac {1} {2}\right)}.
\end{equation}
We can dimensionally regularize  $Z$ - $<U>$ as above, to find

\begin{equation}
\label{eq6.27} 
Z=\frac {V^n} {\Gamma\left(\frac {3 n} {2}\right)}
\left(\frac {3m\pi} {\beta}\right)^{\frac {3 n} {2}}
\frac {(-1)^{\frac {3n-1} {2}}}
{\Gamma\left(\frac {3 n-1} {2}\right)}
\ln\left(\frac {3m\pi} {\beta}\right),
\end{equation}
\begin{equation}
\label{eq6.28} 
<{\cal U}>=
\frac {3} {\beta(3n-1)}
\left[\frac {1} {\ln\beta-\ln 3m\pi}-
\frac {3n} {2}\right].
\end{equation}
We tackle first 
$(-1)^{\frac {3n-1} {2}}=-1$ and then 
$n=1,5,9,13......$
\begin{equation}
\label{eq6.29} 
Z=\frac {V^n}
{\Gamma\left(\frac {3} {2}\right)
\Gamma\left(\frac {3n-1} {2}\right)}
\left(\frac {3m\pi} {\beta}\right)^{\frac {3 n} {2}}
\ln\left(\frac {\beta} {3m\pi}\right).
\end{equation}
For
$(-1)^{\frac {3n-1} {2}}=1$, then  
$n=3,7,11,15......$
\begin{equation}
\label{eq6.30} 
Z=\frac {V^n}
{\Gamma\left(\frac {3} {2}\right)
\Gamma\left(\frac {3n-1} {2}\right)}
\left(\frac {3m\pi} {\beta}\right)^{\frac {3 n} {2}}
\ln\left(\frac {3m\pi} {\beta}\right).
\end{equation}
From  (\ref{eq6.28}) - (\ref{eq6.29}) and demanding e 
$Z>0$ - $<U>>0$ we find
\begin{equation}
\label{eq6.31}
\frac {1} {3m\pi ke^{\frac {2} {3n}}}<T<
\frac {1} {3m\pi k}.
\end{equation}
From  (\ref{eq6.28})-y (\ref{eq6.30}) and requiring  
$Z>0$ - $<U><0$ we get
\begin{equation}
\label{eq6.43}
0\leq T<\frac {1} {3m\pi ke^{\frac {2} {3n}}}.
\end{equation}
Finally, for $C$ one has
\begin{equation}
\label{eq6.44} 
C=
\frac {3k} {3n-1}
\left[\frac {1} {\ln\beta-\ln 3m\pi}+
\frac {1} {(\ln\beta-\ln 3m\pi)^2}-
\frac {3n} {2}\right].
\end{equation}

\section{Conclusions}

\setcounter{equation}{0}

\nd We have used an elementary
regularization procedure to investigate  the poles in both $Z$ and $<U>$ for specific, discrete
q-values, in Tsallis' q-scenario. We analyzed  the thermal
behavior at the poles and encountered suggestive features. The
study was  undertaken  in one, two, three, and $N$ dimensions. Amongst
the ensuing pole-features, rather unexpected, but nonetheless true,  we focus on:

\begin{itemize}

\item There is  an upper bound to the
temperature at the poles, re-confirming the discoveries  of Ref.
\cite{PP94}.

\item In some instances, Tsallis' entropies are positive just for a
restricted temperature-range.

\item Negative specific heats, characteristic feature  of
self-gravitating systems \cite{lb}, are found.
See the illuminating by Silva-Alcaniz \cite{silva}.
We will, in a future work, try to connect our methodology
with that of this reference.

\end{itemize}

Our physical results are deduced only from statistics and 
not from mechanical properties. This fact brings to mind  of a similar
feature that emerges in  the case of the entropic force conjectured by
Verlinde \cite{verlinde}.

\newpage
\begin{figure}[h]
\begin{center}
\includegraphics[scale=0.6,angle=0]{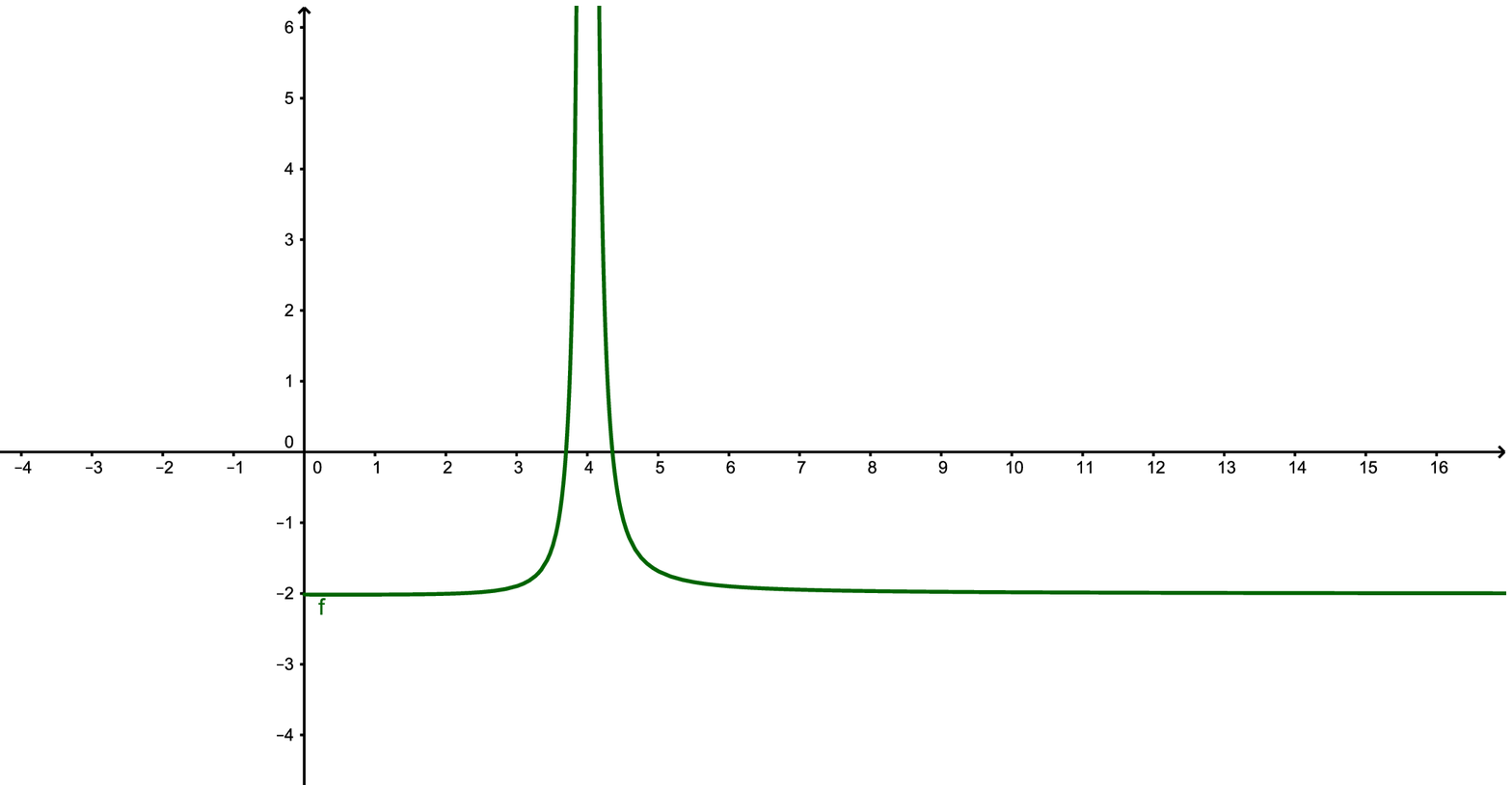}
\vspace{-0.2cm} \caption{$c/k$ at the pole $q=1/2$
versus $\beta$ for $m=1/\pi$ and $n=100$.
The right branch corresponds to $Z>0$, i.e., 
the physical branch.}\label{fig1}
\end{center}
\end{figure}

\newpage
\begin{figure}[h]
\begin{center}
\includegraphics[scale=0.6,angle=0]{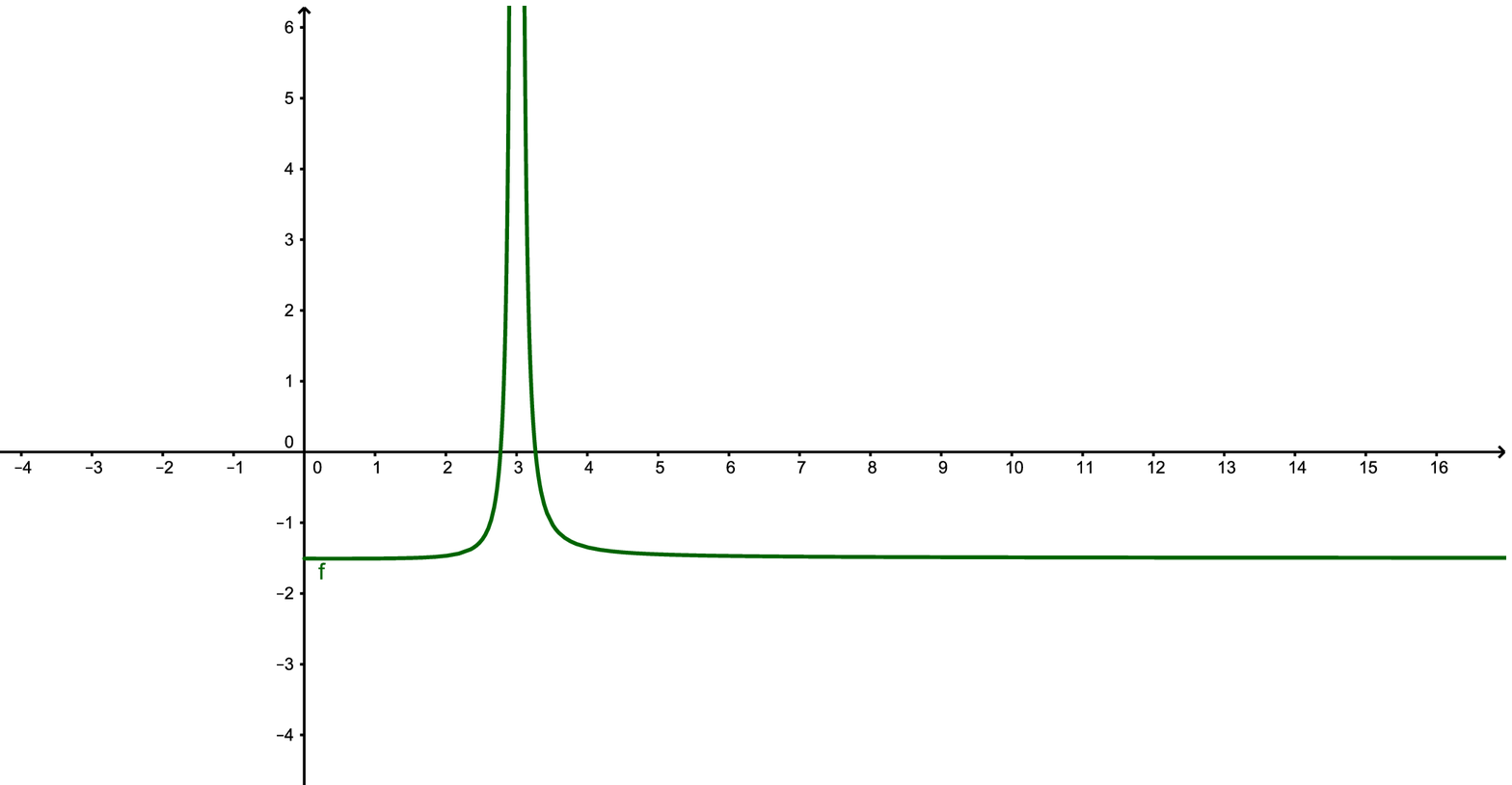}
\vspace{-0.2cm} \caption{$c/k$ at the pole $q=1/3$
versus $\beta$, for $m=1/\pi$ and $n=99$.
The right branch corresponds to $Z>0$, i.e., 
the physical branch.}\label{fig2}
\end{center}
\end{figure}

\end{document}